\documentclass[aps,prd]{revtex4}
\usepackage{graphicx}

\newcommand{\uc}{\uppercase}

\newcommand{\cd}{\! \cdot \!}
\newcommand{\be}{\begin{equation}}
\newcommand{\ee}{\end{equation}}
\newcommand{\ba}{\begin{eqnarray}}
\newcommand{\ea}{\end{eqnarray}}
\renewcommand{\slash}{ \not}
\newcommand{\fermionright}[2]{ \put(#1,#2){\vector(1,0){13}}
\put(#1,#2){\line(1,0){20}} }
\newcommand{\fermionleft}[2]{ \put(#1,#2){\vector(-1,0){13}}
\put(#1,#2){\line(-1,0){20}} }

\begin{document}

\title{Relativistic Quark Model Calculation of the \lowercase{$l_1$,
$l_2$} Coefficients of the Chiral Lagrangian}

\author{Felipe J. Llanes-Estrada
\footnote{
\uc{T}alk  presented by \uc{F}. \uc{L}lanes-\uc{E}strada. \uc{T}hanks
to \uc{E}. \uc{R}ibeiro, \uc{P}. \uc{M}aris, \uc{S}. \uc{C}otanch, \uc{A}.
\uc{S}zczepaniak and \uc{J}. \uc{P}elaez for valuable input.
\uppercase{T}his work supported by an \uppercase{U}niv. 
\uppercase{C}omplutense travel allowance,
and project \uc{FPA} 2000-0956.}}
\address{Depto. Fisica Teorica I, Univ. Complutense, Madrid,
Spain}

\author{Pedro Bicudo}
\address{Dep .F\'{\i}sica and  CFIF, 
Instituto Superior T\'ecnico, 1049-001 Lisboa, Portugal}

\begin{abstract}
We briefly report on a relativistic quark model scheme
to calculate the $O(P^4)$ pion-pion vertex in the planar approximation and
in the chiral limit. The calculation is reduced to the
solution of simple integral equations (Bethe-Salpeter like) by an
effective use of chiral Ward Identities. Specific model computations
are provided.
\end{abstract}
\maketitle

The interest of meson scattering in the quark model has recently revived
due to the exotic $1^{-+}$ mesons seen in BNL at the energies of 
$1.4$ and $1.6 \ GeV$. The hybrid 
\cite{LlanesCotanch} 
or meson molecule nature of these states is being debated.
The standard tools for meson scattering are the chiral
lagrangian 
\cite{GaLe1984} 
and its unitarized extensions 
\cite{Pelaez},
which cannot access that intermediate energy region, let
alone interpret the Fock space nature of the resonances. Still they
provide the benchmark for model approaches to the underlying dynamics
which encompass chiral symmetry.
In recent works 
\cite{weall,Bicudo} 
the use of a novel chiral Ward Identity (\ref{GammaA} below)
allowed the reduction of the planar quark model approximation to pion
scattering,
\begin{equation} \label{PIPISCATT}
\begin{picture}(100,30)(0,0)
\put(0,-5){
\begin{picture}(90,40)(0,0)
\put(25,10){\begin{picture}(40,20)(0,0)
\put(0,15){\line(1,0){5}}
\put(0,5){\vector(1,0){10}}
\put(15,15){\vector(-1,0){10}}
\put(15,5){\line(-1,0){5}}
\put(15,0){\framebox(10,20){}}
\put(25,15){\line(1,0){5}}
\put(25,5){\vector(1,0){10}}
\put(40,15){\vector(-1,0){10}}
\put(40,5){\line(-1,0){5}}
\end{picture}}
\put(25,10){\begin{picture}(40,20)(0,0)
\put(40,10){\oval(30,10)[r]}
\put(42,2){$*$}
\put(48,-4){$\chi_{\pi_3}$}
\put(42,13){$*$}
\put(48,20){$\chi_{\pi_2}$}
\put(55,10){\vector(0,1){2}}
\end{picture}}
\put(-20,10){\begin{picture}(40,20)(0,0)
\put(45,10){\oval(30,10)[l]}
\put(38,2){$*$}
\put(34,-4){$\chi_{\pi_4}$}
\put(38,13){$*$}
\put(34,20){$\chi_{\pi_1}$}
\put(30,10){\vector(0,-1){2}}
\end{picture}}
\end{picture}
}
\end{picture}
+
\begin{picture}(100,30)(0,0)
\put(0,-5){
\begin{picture}(90,40)(0,0)
\put(25,10){\begin{picture}(40,20)(0,0)
\put(0,15){\line(1,0){5}}
\put(0,5){\vector(1,0){10}}
\put(15,15){\vector(-1,0){10}}
\put(15,5){\line(-1,0){5}}
\put(15,0){\framebox(10,20){}}
\put(25,15){\line(1,0){5}}
\put(25,5){\vector(1,0){10}}
\put(40,15){\vector(-1,0){10}}
\put(40,5){\line(-1,0){5}}
\end{picture}}
\put(25,10){\begin{picture}(40,20)(0,0)
\put(40,10){\oval(30,10)[r]}
\put(42,2){$*$}
\put(48,-4){$\chi_{\pi_2}$}
\put(42,13){$*$}
\put(48,20){$\chi_{\pi_1}$}
\put(55,10){\vector(0,1){2}}
\end{picture}}
\put(-20,10){\begin{picture}(40,20)(0,0)
\put(45,10){\oval(30,10)[l]}
\put(38,2){$*$}
\put(34,-4){$\chi_{\pi_3}$}
\put(38,13){$*$}
\put(34,20){$\chi_{\pi_4}$}
\put(30,10){\vector(0,-1){2}}
\end{picture}}
\end{picture}
}
\end{picture}
-
\begin{picture}(65,30)(0,0)
\put(0,-5){
\begin{picture}(65,40)(0,0)
\put(25,10){\begin{picture}(40,20)(0,0)
\put(2,15){\line(1,0){5}}
\put(0,5){\vector(1,0){10}}
\put(15,15){\vector(-1,0){10}}
\put(15,5){\line(-1,0){5}}
\end{picture}}
\put(0,10){\begin{picture}(40,20)(0,0)
\put(40,10){\oval(30,10)[r]}
\put(42,2){$*$}
\put(48,-4){$\chi_{\pi_2}$}
\put(42,13){$*$}
\put(48,20){$\chi_{\pi_1}$}
\put(55,10){\vector(0,1){2}}
\end{picture}}
\put(-20,10){\begin{picture}(40,20)(0,0)
\put(45,10){\oval(30,10)[l]}
\put(38,2){$*$}
\put(34,-4){$\chi_{\pi_3}$}
\put(38,13){$*$}
\put(34,20){$\chi_{\pi_4}$}
\put(30,10){\vector(0,-1){2}}
\end{picture}}
\end{picture} }  
\end{picture}
\end{equation}
to  reproduce the $O(P^2)$, $O(m_\pi^2)$ chiral amplitude. Here we continue this effort 
by establishing contact with the $O(p^4)$, $O(m_\pi^0)$ chiral lagrangian. 
This work is related to microscopic studies of $\pi-\pi$ scattering
with the Nambu and Jona-Lasinio model
\cite{Brigitte},
with the bosonization technique
\cite{Craig},
with the equal time quark model 
\cite{weall,Goncalo},
and with a direct numerical evaluation with a Schwinger-Dyson model
\cite{weall,Steve}. While the $O(P^2)$ amplitude is model independent, 
the $O(p^4)$ provides a model test.  
Here we exactly and explicitly 
determine the Feynman diagrams that govern the $O(p^4)$ amplitude, then
employ two simple models for a test evaluation.

In the standard notation, the pion scattering amplitude arising from its 
$O(P^4)$ contact terms alone reads
\be \label{defthels}
A(s,t,u)^{(P^4)}=\frac{1}{F^4} [(2 l_1 + \frac{l_2}{2})s^2 + \frac{l_2}{2} 
(t-u)^2]\ .
\ee
The Bethe-Salpeter pion vertex $\chi_\pi$ in (\ref{PIPISCATT}) above, function
of two momenta (incoming meson $P$ and quark $k+P/2$) can be
conveniently expanded in an external momentum power series. But first it
is useful to extract from it a dressed vertex $\Gamma_A$ 
satisfying interesting Ward Identities connecting it to the fermion dressed
propagator and the planar ladder,
\ba \label{GammaA}
\chi (P,k) &=& \frac{-i \Gamma_A(P,k) + \Delta(P,k)}{\sqrt{2}f_\pi}
\\ \nonumber
\Gamma_A(P,k+P/2)&=&S^{-1}(k-P/2)\gamma_5+\gamma_5 S^{-1}(k+P/2)
\\ \nonumber
\begin{picture}(90,35)(0,5)
\put(5,10){\begin{picture}(40,20)(0,0)
\put(0,15){\line(1,0){5}}
\put(0,5){\vector(1,0){10}}
\put(15,15){\vector(-1,0){10}}
\put(15,5){\line(-1,0){5}}
\put(15,0){\framebox(10,20){}}
\put(25,15){\line(1,0){5}}
\put(25,5){\vector(1,0){10}}
\put(40,15){\vector(-1,0){10}}
\put(40,5){\line(-1,0){5}}
\end{picture}}
\put(5,10){\begin{picture}(60,20)(0,0)
\put(38,2){$*$}
\put(35,-8){$S^{-1}$}
\put(38,13){$*$}
\put(37,20){$\Gamma_{\hspace{-.08cm}A}$}
\end{picture}}
%
\put(45,10){\begin{picture}(40,20)(0,0)
\put(0,15){\line(1,0){5}}
\put(0,5){\vector(1,0){10}}
\put(15,15){\vector(-1,0){10}}
\put(15,5){\line(-1,0){5}}
\put(15,0){\framebox(10,20){}}
\put(25,15){\line(1,0){5}}
\put(25,5){\vector(1,0){10}}
\put(40,15){\vector(-1,0){10}}
\put(40,5){\line(-1,0){5}}
\end{picture}}
%
\end{picture}
&=&
\begin{picture}(105,35)(0,5)
\put(15,5){\begin{picture}(40,20)(0,0)
\put(-5,13){$\gamma_5$}
\put(10,15){\line(1,0){5}}
\put(10,5){\vector(1,0){10}}
\put(25,15){\vector(-1,0){10}}
\put(25,5){\line(-1,0){5}}
\put(25,0){\framebox(10,20){}}
\put(35,15){\line(1,0){5}}
\put(35,5){\vector(1,0){10}}
\put(50,15){\vector(-1,0){10}}
\put(50,5){\line(-1,0){5}}
\end{picture}}
\end{picture}
+
\begin{picture}(75,35)(0,5)
\put(5,5){\begin{picture}(40,20)(0,0)
\put(0,15){\line(1,0){5}}
\put(0,5){\vector(1,0){10}}
\put(15,15){\vector(-1,0){10}}
\put(15,5){\line(-1,0){5}}
\put(15,0){\framebox(10,20){}}
\put(25,15){\line(1,0){5}}
\put(25,5){\vector(1,0){10}}
\put(40,15){\vector(-1,0){10}}
\put(40,5){\line(-1,0){5}}
\put(45,13){$\gamma_5$}
\end{picture}}
\end{picture}
\ .
\ea
\newpage
The use of these identities, as well as the fact that no zero-momentum 
pole appears in any vector or scalar ladder given the symmetry breaking 
pattern of the theory, allows us to reduce the momentum four term of 
eq. (\ref{PIPISCATT}) to three different classes of diagrams, 
parameterized as,
\ba \label{Defined12}
\begin{picture}(100,35)(0,0) 
\put(0,-5){
\begin{picture}(100,40)(0,0)
\put(25,10){\begin{picture}(40,20)(0,0)
\put(0,15){\line(1,0){5}}
\put(0,5){\vector(1,0){10}}
\put(15,15){\vector(-1,0){10}}
\put(15,5){\line(-1,0){5}}
\put(15,0){\framebox(10,20){}}
\put(25,15){\line(1,0){5}}
\put(25,5){\vector(1,0){10}}
\put(40,15){\vector(-1,0){10}}
\put(40,5){\line(-1,0){5}}
\end{picture}}
\put(25,10){\begin{picture}(40,20)(0,0)
\put(40,10){\oval(30,10)[r]}
\put(42,2){$*$}
\put(48,-4){$\Delta_3$}
\put(42,13){$*$}
\put(48,20){$\Delta_2$}
\put(55,10){\vector(0,1){2}}
\end{picture}}
\put(-20,10){\begin{picture}(40,20)(0,0)
\put(45,10){\oval(30,10)[l]}
\put(38,2){$*$}
\put(34,-4){$\Delta_4$}
\put(38,13){$*$}
\put(34,20){$\Delta_1$}
\put(30,10){\vector(0,-1){2}}
\end{picture}}
\end{picture} } 
\end{picture}
+
\begin{picture}(100,35)(0,0) 
\put(0,-5){
\begin{picture}(100,40)(0,0)
\put(25,10){\begin{picture}(40,20)(0,0)
\put(0,15){\line(1,0){5}}
\put(0,5){\vector(1,0){10}}
\put(15,15){\vector(-1,0){10}}
\put(15,5){\line(-1,0){5}}
\put(15,0){\framebox(10,20){}}
\put(25,15){\line(1,0){5}}
\put(25,5){\vector(1,0){10}}
\put(40,15){\vector(-1,0){10}}
\put(40,5){\line(-1,0){5}}
\end{picture}}
\put(25,10){\begin{picture}(40,20)(0,0)
\put(40,10){\oval(30,10)[r]}
\put(42,2){$*$}
\put(48,-4){$\Delta_2$}
\put(42,13){$*$}
\put(48,20){$\Delta_1$}
\put(55,10){\vector(0,1){2}}
\end{picture}}
\put(-20,10){\begin{picture}(40,20)(0,0)
\put(45,10){\oval(30,10)[l]}
\put(38,2){$*$}
\put(34,-4){$\Delta_3$}
\put(38,13){$*$}
\put(34,20){$\Delta_4$}
\put(30,10){\vector(0,-1){2}}
\end{picture}}
\end{picture} } 
\end{picture}
-
\begin{picture}(65,35)(0,0) 
\put(0,-5){
\begin{picture}(65,40)(0,0)
\put(25,10){\begin{picture}(40,20)(0,0)
\put(0,15){\line(1,0){5}}
\put(0,5){\vector(1,0){10}}
\put(15,15){\vector(-1,0){10}}
\put(15,5){\line(-1,0){5}}
\end{picture}}
\put(0,10){\begin{picture}(40,20)(0,0)
\put(40,10){\oval(30,10)[r]}
\put(42,2){$*$}
\put(48,-4){$\Delta_2$}
\put(42,13){$*$}
\put(48,20){$\Delta_1$}
\put(55,10){\vector(0,1){2}}
\end{picture}}
\put(-20,10){\begin{picture}(40,20)(0,0)
\put(45,10){\oval(30,10)[l]}
\put(38,2){$*$}
\put(34,-4){$\Delta_3$}
\put(38,13){$*$}
\put(34,20){$\Delta_4$}
\put(30,10){\vector(0,-1){2}}
\end{picture}}
\end{picture} } 
\end{picture}
\\ 
= 3 d_1 (P_1\cd P_2 \ P_3 \cd P_4+ P_1
\cd P_4 \ P_2 \cd P_3) + 3 d_2 P_1 \cd P_3 \ P_2 \cd P_4  \ ,
\nonumber \\
\begin{picture}(250,40)(0,5)
\put(0,22){$\slash{P}_3$}
\put(20,25){\oval(10,10)[l]} \put(12,23){$*$}
\fermionleft{40}{30} \fermionright{20}{20}
\put(20,35){k}
\put(15,7){$k\! \!-\! \! P_1\! \! - \! \! P_2$}
\put(40,16){\framebox(10,18){}}
\fermionleft{70}{30} \fermionright{50}{20}
\put(70,25){\oval(10,10)[r]}
\put(70,27){$*$} \put(75,35){$\Delta_1$}
\put(70,18){$*$} \put(75,8){$\Delta_2$}
\put(90,18){
$=3 d_3 P_1\cd P_3 \ P_1 \cd P_2 + 3 d_4 P_2 \cd P_3 \ P_1 \cd P_2$ ,}
\end{picture}
\nonumber \\
\begin{picture}(260,40)(0,5)
\put(0,22){$\slash{P}_1$}
\put(20,25){\oval(10,10)[l]} \put(12,23){$*$}
\fermionleft{40}{30} \fermionright{20}{20}
\put(40,16){\framebox(10,18){}}
\fermionleft{70}{30} \fermionright{50}{20}
\put(70,25){\oval(10,10)[r]}
\put(73,23){$*$} \put(75,22){$\slash{P}_3$}
\put(25,32){$q$} \put(17,5){$^{q\! + \! P_1 \! +\! P_4}$}
\put(95,18){
$= 3 d_5 P_1 \cd P_3 \ P_1 \cd P_4 + 3 d_6 P_3 \cd P_4 \ P_1 \cd P_4$ .}
\nonumber
\end{picture}
\ea 
All external momentum and color dependence is explicit. 
To calculate the ladders which remain in eq. (\ref{Defined12}), 
we write integral equations for them, analogous to the Bethe-Salpeter 
equation for a meson and with up to ten independent functions of $q^2$. 
Matching this Feynman amplitude with the physical external momenta and 
multiplying each permutation by the adequate isospin factors, we obtain 
for the $l$'s defined in eq. (\ref{defthels}), 
\ba 
l_1&=& { 3 \over 32}(-2 d_1 +  d_2 - 6 d_3 + 2 d_4 + 5 d_5 +  d_6)\ 
\nonumber \\
l_2&=& -{3 \over 16} (d_2 - 2d_3 + 2d_4 +d_5 -d_6)
\ . 
\ea 
The necessary Dirac gamma traces are computed
with FORM 
\cite{Vermasseren}. 
We perform model evaluations in Euclidean space with an interaction rung 
given by $V=-g^2 \gamma_\mu i K(k-q)  \gamma^\mu$ which is similar to the
One Gluon Exchange interaction in the Feynman Gauge. We have used two
different infrared strong kernels, both UV finite to avoid
renormalization complications.

The first is a simple Gaussian kernel (whose Euclidean angular
integrals are simply Bessel functions)
$ K(q)=exp(-q^2/\Lambda^2)$
with parameters $g=3$, $\Lambda=1.15 \ GeV$ (model 1) which leads to a
constituent quark mass of $116 \ MeV$, a condensate $-(170 \
MeV)^3$, a pion decay constant $f_\pi=87 \ MeV$ and gives 
$l_1=-0.11$, $l_2= 0.14$; while the parameters (model 2)
 $g=1.7$, $\Lambda=2$ lead to a mass of $93 \ MeV$, condensate  $-(229MeV)^3$,
and $f_\pi=86 \ MeV$, $l_1= -0.06$, $l_2=0.12$.
The second is a rational kernel (whose Euclidean angular
integrals contain only polynomials and square roots)
$ K(q)=[ 1/(q^2-\lambda^2) -1/(q^2-\Lambda^2)] $
with $g=10.3$, $\lambda=0.8 \ GeV$, $\Lambda = 0.85 \ GeV$ (model 3), we
obtain a constituent quark mass of $216 \ MeV$, condensate $-(156
\ MeV )^3$, pion decay constant $92 \ MeV$, and $l_1=-0.065$,
$l_2=-0.12$. 

The computed $l_1$ and $l_2$ have the correct signs and ratio
when compared with the phenomenological analysis, however they
are large. Investigations are underway to test a wider class
of quark-antiquark interactions.

\begin{figure}
\includegraphics[scale=.76]{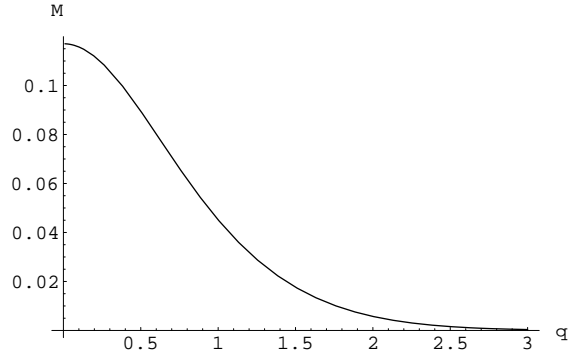}
\vspace{-.3cm}
\caption{\rm \label{fig1}  Mass gap function solution of the Schwinger-Dyson 
equation for model 1 (in GeV).}
\end{figure}

\begin{figure}
\includegraphics[scale=.76]{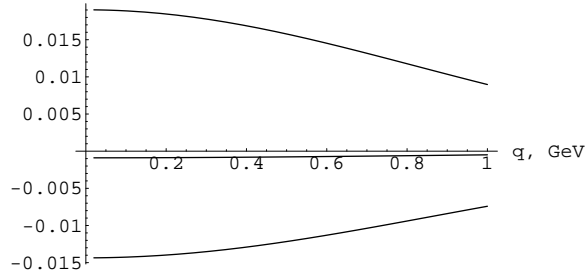}
\vspace{-.3cm}
\caption{\rm \label{fig2}  Solution of the homogeneous Bethe-Salpeter equation. From top 
to bottom, $F0[a]$, $E2[q]$, $G0[q]$ (corresponding to the coefficients of 
an external momentum expansion of the 
llewelyn-Smith wavefunctions). }
\end{figure}

\begin{figure} 
\includegraphics[scale=.76]{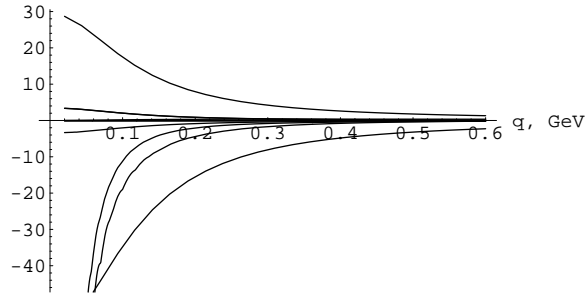}
\vspace{-.3cm}
\caption{\rm \label{fig3}  Solution of the homogeneous equation for two $\Delta$'s with ladder,
needed to calculate d1,d2,d3,d4. (Coefficients of external  momentum expansion).}
\end{figure}

\begin{figure} 
\includegraphics[scale=.76]{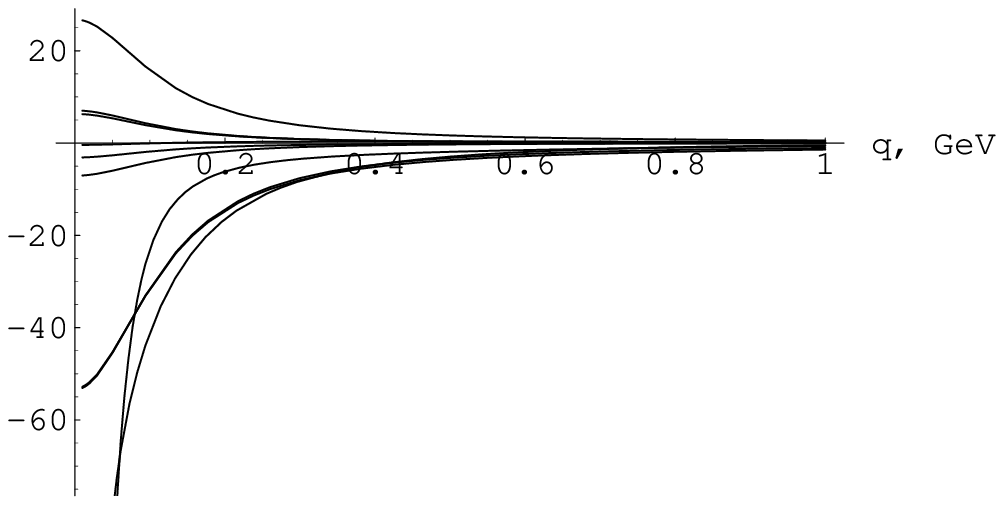}
\vspace{-.3cm}
\caption{\rm \label{fig4}  Solution of the homogeneous equation for $\slash{P}$ with ladder,
needed to calculate d5,d6. (Coefficients of external momentum expansion).}
\end{figure}

\end{document}